\documentclass[12pt]{article}
\usepackage[numbers,sort&compress]{natbib}

\usepackage[utf8]{inputenc}
\usepackage[T1]{fontenc}
\usepackage{amssymb, amsmath, mathrsfs}   
\usepackage{graphicx}
\usepackage{leading}   

\usepackage[margin=2cm]{geometry}
\usepackage[colorlinks=true, allcolors=blue]{hyperref}
\usepackage{orcidlink}

\title{\bf\Large Scale-dependent gravity and covariant scale-setting}
\author{{\bf \normalsize Nicolas R. Bertini} \orcidlink{0000-0003-0301-0610}\footnote{nicolas.bertini@hotmail.com}\\[15pt]
	\textit{Departamento de Física, ICE, Universidade Federal de Juiz de Fora}\\
	\textit{Campus Universitário - Juiz de Fora, 36036-900, MG, Brazil}}
\date{}
	
\begin{document}
	
\maketitle
\leading{18pt}

\vspace{-.4cm}
\begin{abstract}
	\noindent
	A fundamental element of scale-dependent gravity is the scale-setting procedure. We present a new covariant expression to set the scale that arises when examining the field equations. Considering the renormalization group equations and imposing energy-momentum tensor conservation, we arrive at two models of running of the gravitational and cosmological constants. In the cosmological setting,  we found that in one model the Big Bang singularity is avoided, while in the other the Hubble tension can be alleviated. At the level of cosmological perturbations, we derived the basic solutions and qualitatively discussed the impacts of this scenario on structure formation.
\end{abstract}

\section{Introduction}
	
Despite explaining most of the observational data on astrophysical and cosmological scales, General relativity (GR) presents problems in the theoretical and phenomenological sectors, such as non-renormalizability at the quantum level, the prediction of singularities, the inability to fully explain the nature of the dark sector, and tensions involving some cosmological parameters. In an attempt to address these limitations and considering that GR can be modified in various directions, the search for alternative theories has become a fruitful field of investigation. This has resulted in a wide variety of models available in the literature \cite{Clifton:2011jh}. On the other hand, it is important to reconsider whether one could contribute to the understanding of the aforementioned problems, using the well-established physical framework. 
In this regard, quantum field theory in curved spacetime has been considered to extend GR across different energy scales, incorporating effects from the renormalization group (RG). This includes semiclassical approaches, which emphasize observational constraints and classical symmetries \cite{Shapiro:2004ch,Rodrigues:2015hba, Bertini:2019xws}, as well as approaches that account for the effects of quantized gravity, either in perturbative formulations \cite{Donoghue:1994dn, Giacchini:2020zrl, Giacchini:2020dhv} or non-perturbative frameworks, such as asymptotic safety in quantum gravity \cite{ Reuter:1996cp, Bonanno:2000ep, Percacci:2007sz}.

In the context of quantum gravity, higher derivative terms must be added to the Einstein-Hilbert action to ensure a consistent quantization of the matter sector. These terms are dynamically relevant at short distances but become negligible at larger scales. Moreover, it can be shown that the running couplings associated with these terms have trivial RG flows in the infrared, meaning they become genuine constants \cite{Gorbar:2002pw, Gorbar:2003yt}. This behavior is analogous to quantum electrodynamics, where the coupling approaches a true constant in the infrared limit (see, e.g., \cite{Goncalves:2009sk, KLOE-2:2016mgi}). However, the gravitational and cosmological constants, $G$ and $\Lambda$, do not necessarily follow the same pattern and may run in the deep infrared (see, e.g., \cite{Shapiro:2009dh}). Although they can run, is not defined their precise running behavior each context. The two key unknowns are the $\beta$-functions, which govern the scale dependence of the running couplings, and the relation between the renormalization scale $\mu$ and other physical quantities. This last connection is known as the scale-setting procedure (see, e.g., \cite{Reuter:2003ca, Shapiro:2004ch, Babic:2004ev, Domazet:2010bk, Koch:2014joa}).

Here, we consider a class of corrections to GR based on the expectations related to the RG at large distance scales. In this scenario, $G$ and $\Lambda$ are viewed as running couplings that depend on $\mu$ and are constructed by integrating the corresponding $\beta$-functions. This scale-dependent approach has shown promising in understanding the evolution of the universe \cite{Shapiro:1999zt, Shapiro:2000dz, Babic:2001vv, Guberina:2002wt, Shapiro:2003ui, Espana-Bonet:2003qjh} and has several phenomenologically interesting applications. For instance, in the cosmological dark sector \cite{Sola:2005et, Sola:2005nh}, the coincidence problem \cite{Grande:2006nn, Grande:2006qi}, the growth of inhomogeneities \cite{Fabris:2006gt, Grande:2007wj, Grande:2008re, Grande:2010vg}, the relaxation mechanism for the cosmological term \cite{Bauer:2009ke, Bauer:2009jk}, and addressing mechanisms that can alleviate the Hubble tension \cite{Alvarez:2020xmk, Bertini:2024onw}.
	
This scale dependence in GR has been considered primarily on two levels: in the field equations or in the action. Regardless of the approach, any proposal in this direction must address the two key points aforementioned: $i$) the relationships between $G$ and $\Lambda$ and the scale $\mu$, or equivalently, provide the corresponding $\beta$-functions; and $ii$) the relationship between $\mu$ and other relevant physical quantities. The second point is typically addressed within a specific physical context. In background cosmology, the most well-motivated identification is $\mu = H$, where $H$ is the Hubble parameter \cite{Shapiro:2000dz, Shapiro:2004ch}, whereas for systems at astrophysical scales, $\mu \propto r^{-1}$, where $r$ is the radial distance from the system’s center \cite{Shapiro:2004ch, Domazet:2010bk}. This functional dependence on coordinates is specific to the chosen coordinate system and depends on the running model, i.e., the explicit forms of $G(\mu)$ and $\Lambda(\mu)$ \cite{Domazet:2010bk}. In the physical situations where an expression independent of coordinates is suitable, such as developing cosmological perturbations, an ad hoc expression is often introduced to recover the background relation \cite{Fabris:2006gt}, or that works specifically within the astrophysical context (see, e.g., \cite{Toniato:2017wmk, Bertini:2019xws} and references therein).

It is important to note that most scale-identification procedures are tied to specific physical situations, meaning they are not covariant. However, covariance is a fundamental feature of the effective action of vacuum, which is assumed to underlie the running of couplings. In this paper, we propose a new method to identify \( \mu \) with a covariant expression derived from the field equations, without relying on the specific forms of \( G(\mu) \) and \( \Lambda(\mu) \). We considered implementing scale dependence at the level of the equations of motion, but this identification need not be restricted to this approach. 
For practical purposes, one can follow the common strategy (e.g., \cite{Bonanno:2001xi, Shapiro:2003ui, Shapiro:2004ch, Rodrigues:2009vf, Perico:2013mna, Arboleda:2018qdo, Bertini:2019xws} and references therein) and assume energy-momentum conservation. In this context, the Bianchi identities imply a consistency equation widely used both in the RG-based scale-setting procedure and to determine a pair \( \{G(\mu),\Lambda(\mu)\} \) compatible with the conservation law (see, e.g., \cite{Babic:2004ev, Shapiro:2004ch, Domazet:2010bk}); one running coupling is obtained via the RG equations, while the other follows from the consistency relation, given a specific physical system and an appropriate scale-setting prescription.  
The procedure adopted here does not depend on energy-momentum tensor conservation. However, if this assumption is made, we derive a relation between the couplings that can be combined with the RG equations to obtain viable expressions for \( G(\mu) \) and \( \Lambda(\mu) \), without the need to specify a particular physical system or coordinate choice.  

To exemplify the consequences of this framework, we applied it to background cosmology and perturbations, considering two running coupling models derived from the RG $\beta$-functions. The first model arises from the infrared $\beta$-function for $G$, leading to a logarithmic dependence of this coupling with $\mu$. This behavior has been obtained in several works that include semiclassical approaches \cite{Buchbinder:1992rb, BruceL:1982}, theories with higher derivatives \cite{Fradkin:1981iu, Modesto:2017hzl} and, as argued in \cite{Farina:2011me}, is expected to arise in any renormalizable theory employing the minimal subtraction scheme. The second model, in contrast, highlights a scenario with significant implications at higher energy scales. In this case, the correction to the Friedmann equation manifests as an additional term proportional to $H^{4}$, providing an attractive alternative to the singular description of the primordial universe obtained through standard GR. These models are investigated with a focus on analytical results and the qualitative understanding of the implied cosmology.

This paper is organized as follows. In the next Sec.~\ref{sec:2}, the field equations and the scale identification process are presented. In Sec. \ref{sec:models} we derive the running of the gravitational constant and cosmological term, ensuring compatibility with the conservation law of the energy-momentum tensor. Analytical expressions for the Hubble parameter are obtained and discussed in Sec.~\ref{sec:back_cosmo}. Cosmological perturbations are developed in Sec.~\ref{sec:pert}. Conclusions and perspectives are presented in Sec.~\ref{sec:conclusions}, and appendix \ref{sec:AP1} shows the correspondence $\mu\propto r^{-1}$ when considering a static and spherically symmetric metric. We adopt the following notations and conventions: the Riemann tensor is defined as $R^{\alpha}_{\,\,\, \beta\sigma\rho} = \partial_{\sigma}\Gamma^{\alpha}_{\beta\rho} - \partial_{\rho}\Gamma^{\alpha}_{\beta\sigma}+\Gamma^{\alpha}_{\lambda\sigma}\Gamma^{\lambda}_{\beta\rho}- \Gamma^{\alpha}_{\lambda\rho}\Gamma^{\lambda}_{\beta\sigma}$, the Ricci tensor as $R_{\beta\rho}=R^{\alpha}_{\, \,\,\beta\alpha\rho}$ and the Ricci scalar as $R=g^{\alpha\beta}R_{\alpha\beta}$. The metric signature is taken to be $(+,-,-,-)$.

\section{Viable running couplings and covariant scale-setting}\label{sec:2}

As previously mentioned, although the fundamental action of gravity may contain higher-derivative terms, these contributions are generally suppressed at low energies. Therefore, GR is often viewed as an effective theory valid at certain energy scales. In this sense, since quantum corrections modify the couplings but may not introduce new dynamical degrees of freedom at all scales, the structure of Einstein’s equations may remain valid. The scale dependence of $G$ and $\Lambda$ in this case represents small corrections to the dynamics, effectively capturing the RG effects, and may be sufficient to explain phenomena at astrophysical and cosmological scales. 

Taking these arguments into account and considering that $G$ and $\Lambda$ are functions of a given scale $\mu$, the modified Einstein equations are
\begin{align}
G_{\alpha\beta} = 8\pi G(\mu)\big[T_{\alpha\beta} + g_{\alpha\beta}\rho_\Lambda(\mu)\big],\label{eq:FE}
\end{align}
where $G_{\alpha\beta} = R_{\alpha\beta}-g_{\alpha\beta}R/2$ is the Einstein tensor, $\rho_\Lambda \equiv \Lambda/8\pi G$ and $T_{\alpha\beta}$ is the energy-momentum tensor of the matter fields. Let us assume that $T_{\alpha\beta}$ represents a perfect fluid, i.e.
\begin{align}
T_{\alpha\beta} = (\rho+p)u_{\alpha}u_{\beta} - g_{\alpha\beta}p\,, \label{eq:EMT}
\end{align}
where $p$, $\rho$ and $u^{\alpha}$ are the pressure, energy density and the 4-velocity of the fluid, respectively. The conservation law for such a fluid can be expressed by the continuity equation
\begin{align}
u^{\beta}\nabla_{\alpha}T^{\alpha}_{\,\,\,\,\beta} = u^{\alpha}\nabla_{\alpha}\rho + (\rho+p)\nabla_{\alpha}u^{\alpha}=0\,. \label{eq:cont}
\end{align}
Therefore, considering Eq.~\eqref{eq:FE}, the viable running of $G$ and $\rho_\Lambda$, as well as the scale setting, must be compatible with this restriction. Then, let us derive the continuity equation directly from \eqref{eq:FE} and proceed with the appropriate coordinate-independent scale setting.

The Bianchi identities directly imply
\begin{align}
\nabla_{\alpha}T^{\alpha}_{\,\,\,\,\beta} = - \nabla_{\beta}\rho_\Lambda  - \frac{1}{G}(T^{\alpha}_{\,\,\,\,\beta} + \delta^{\alpha}_{\beta}\rho_\Lambda)\nabla_{\alpha}G\,.
\end{align}
Contracting with $u^{\beta}$ and using $\nabla_{\alpha}\rho_\Lambda = (\nabla_{\alpha}\mu )d\rho_\Lambda/d\mu$ and $\nabla_{\alpha}G = (\nabla_{\alpha}\mu) dG/d\mu$, the above equation becomes
\begin{align}
u^{\beta}\nabla_{\alpha}T^{\alpha}_{\,\,\,\,\beta} = - \left( \frac{d\rho_\Lambda}{d\mu} + \frac{\rho+\rho_\Lambda}{G}\frac{dG}{d\mu} \right) u^{\beta}\nabla_{\beta}\mu\,. \label{eq:div1}
\end{align}
	
This expression, together with the conservation-law based constraint $u^{\beta}\nabla_{\alpha}T^{\alpha}_{\,\,\,\,\beta} = 0$, provides the scale-setting condition for extended GR via RG \cite{Babic:2004ev, Shapiro:2004ch, Domazet:2010bk}.
Since $\rho_\Lambda$ and $G$ can be directly defined in terms of $\mu$ through RG flows, for each system (e.g. in cosmology or astrophysics) this equation establishes $\rho = f(\mu)$. When it is possible to invert this relationship, in the form $\mu = f^{-1}(\rho)$, the scale setting is complete. In what follows, we describe a different approach to solving this problem. 
	
Note that from Eq.~\eqref{eq:FE}, we have
\begin{align}
\rho+\rho_\Lambda = \frac{1}{8\pi G}u^{\alpha}u^{\beta}G_{\alpha\beta}\,.\label{eq:rho}
\end{align}
This means that regardless of the specific forms for $G(\mu)$ e $\rho_{\Lambda}(\mu)$ the connection between $\mu$ and the physical information contained in $\rho$ should relate to $u^{\alpha}u^{\beta}G_{\alpha\beta}$. Thus, one can set
\begin{align}
\mu^{2} = \frac{1}{3}u^{\alpha}u^{\beta}G_{\alpha\beta}\,.\label{eq:ss}
\end{align}
The identification with $\mu^{2}$ is carried out to ensure the correct dimensionality and, together with the factor 1/3, maintain proper identification in the contexts mentioned in the introduction. This identification is also convenient because, as can be seen in the next section, $\mu^{2}$ or $\mu^{4}$ appears in the concrete realizations for $G(\mu)$ and $\rho_\Lambda(\mu)$.
	
Plugging \eqref{eq:ss} into Eq.~\eqref{eq:div1}, it follows that any viable running couplings must satisfy
\begin{align}
\frac{d\rho_\Lambda}{d\mu} - \frac{3\mu^2}{8\pi}\frac{dG^{-1}}{d\mu} = 0\,.\label{eq:viablerun} 
\end{align}
In other words, any realizations for $G(\mu)$ and $\rho_{\Lambda}(\mu)$ that are compatible with the conservation law \eqref{eq:cont} must satisfy the above relation.

\section{Running coupling models}\label{sec:models}
	
To provide examples of running couplings, we examine realizations derived from RG predictions for $G(\mu)$ and $\rho_\Lambda(\mu)$ within the framework of quantum field theory in curved spacetime following a similar approach to Ref.~\cite{Shapiro:2004ch} (and references therein).
Based on the scale dependence of the vacuum energy \cite{Shapiro:1999zt}, and searching for a gravity with variable $G$, one can introduce the following RG equations
\begin{align}
\mu\frac{d\rho_\Lambda}{d\mu} = \sum_{n=0}^{\infty}A_n \mu^{2n}, \qquad \mu\frac{d}{d\mu}\left( \frac{1}{G} \right) = \sum_{n=0}^{\infty}B_n \mu^{2n}.\label{eq:betaf} 
\end{align}
The coefficients $A_n$ and $B_n$ are introduced to account for dimensionality and depend on the contributions from the masses of different fields (for more details see \cite{Shapiro:2003ui, Shapiro:2004ch}). 
	
The right-hand side of the above equations defines the respective $\beta$-functions for $\rho_\Lambda$ and $G$. However, requiring viable running couplings implies that $\rho_\Lambda (\mu)$ and $G(\mu)$ cannot be independent; if one of the expansions is provided, the other must be obtained by solving Eq.~\eqref{eq:viablerun}. In what follows, we choose to provide the $\beta$-functions for $G$ and explore two possible realizations.
	
The infrared $\beta$-function of $G$  corresponds to $B_{n\geq 1}=0$ and $B_{0} = 2\nu M^{2}_{\mathrm{Planck}} = 2\nu G^{-1}_{0}$ \cite{Fradkin:1981iu, BruceL:1982, Reuter:2003ca, Bauer:2005rpa} (see also \cite{Farina:2011me} for well-motivated dimensional arguments), where $\nu$ is a small dimensionless constant. Then the RG equation for $G^{-1}$ is
\begin{align}
\mu\frac{d}{d\mu}\left( \frac{1}{G} \right) = 2\nu G_{0}^{-1}\,. \label{eq:RGEG1}
\end{align}
Integrating \eqref{eq:RGEG1} and \eqref{eq:viablerun} we get
\begin{align}
G(\mu) = \frac{G_0}{1+\nu \ln(\mu^{2}/\bar{\mu}^{2})},\qquad \rho_\Lambda(\mu) = \rho_{\Lambda 0} + \frac{3\nu}{8\pi G_0}(\mu^{2}-\bar{\mu}^{2}),\label{eq:ss_quadratic}
\end{align}
where $\bar\mu$ represents a reference scale such that
\begin{align}
G(\bar{\mu}) = G_0, \qquad  \rho_{\Lambda}(\bar\mu)=\rho_{\Lambda 0},\label{eq:boundary}
\end{align}
with $\rho_{\Lambda 0} \equiv \Lambda_0/8\pi G_{0}$, where $\Lambda_0$ the current value of the cosmological constant.
	
Another possibility is to assume that the main contribution comes from the second term in the expansion to $G^{-1}$ in \eqref{eq:betaf}, i.e., $B_{n\neq 1}=0$. Writing $B_1 = 2\zeta$, we have
\begin{align}
\frac{d}{d\mu}\left( \frac{1}{G} \right) = 2
\zeta \mu\label{eq:RGEG2}\,.
\end{align}
Integrating Eqs.~\eqref{eq:RGEG2} and \eqref{eq:viablerun} under the boundary conditions \eqref{eq:boundary}, one gets
\begin{align}
G(\mu) = \frac{G_0}{1+\zeta G_0 (\mu^{2} - \bar{\mu}^{2})}, \qquad \rho_{\Lambda}(\mu) = \rho_{\Lambda 0} + \frac{3\zeta}{16\pi}(\mu^{4} - \bar{\mu}^{4}).\label{eq:ss_quartic1}
\end{align}

It is worth emphasizing that the running laws \eqref{eq:ss_quadratic} and \eqref{eq:ss_quartic1} can also be equivalently derived by providing the $\beta$-function for $\rho_\Lambda$ and integrating the expansion \eqref{eq:betaf} for $\rho_\Lambda$ together with \eqref{eq:viablerun} under the boundary conditions \eqref{eq:boundary}. This approach is adopted in Ref. \cite{Shapiro:2004ch} using a specific coordinate system suitable for background cosmology, under the assumption $\mu = H$. In this context, due to the smallness of $G_0 H^{2}$ and $H^{4}$, the running effects of the couplings in \eqref{eq:ss_quartic1} are expected to become relevant at higher energy scales. To illustrate, consider the integrated version of the expansion for $\rho_\Lambda $ in \eqref{eq:betaf}, expressed as $\rho_\Lambda(H) = \rho_{\Lambda 0} + \frac{3\nu}{8\pi G_0} (H^{2} - \bar{\mu}^{2}) + {\cal O}(H^{2})$. Notably, if a symmetry imposes $\nu = 0$ --- a scenario that may arise at ultraviolet scales \cite{Shapiro:2008yu, Shapiro:2000dz} --- the next dominant term is $\propto H^{4}$. Consequently, requiring the conservation of the energy-momentum tensor naturally leads to a quadratic dependence of \( G(H) \). Indeed, as shown in the next section, these effects directly influence the description of the primordial universe.

In contrast, the approach developed here removes the reliance on a specific coordinate system, thereby broadening the scope and applicability of above running laws.
	
\section{Background cosmology}\label{sec:back_cosmo}
	
In this section we derive the solutions to the background cosmology equations. We assume a homogeneous and isotropic universe, given by the Friedmann-Lemaître-Robertson-Walker (FLRW) metric
\begin{align}
ds^{2} = dt^2 - a^{2}(t)\delta_{ij}dx^idx^j\,,\label{eq:metricfried}
\end{align}
where $a(t)$ is the cosmological scale factor. Adopting a comoving frame, the identification \eqref{eq:ss} implies $\mu = H$, where $H\equiv \dot{a}/a$. Therefore, the 0-0 and i-j components of the modified Einstein equations \eqref{eq:FE}, are
\begin{align}
3H^{2} & = 8\pi G(H)\big[\rho+\rho_\Lambda(H)\big],\label{eq:fried1}
\end{align}
\begin{align}
2\dot{H} + 3H^{2} & = - 8\pi G(H)\big[ p -\rho_{\Lambda}(H)\big].\label{eq:fried2}
\end{align}
Additionally, the conservation law \eqref{eq:cont} implies
\begin{align}
\dot{\rho}+ 3H(\rho+p)=0\,. \label{eq:cont_back}
\end{align}
We will now consider $G$ and $\rho_\Lambda$ presented in the previous section.

\subsection{Logarithmic running of \texorpdfstring{$G$}{G} and quadratic running of \texorpdfstring{$\rho_\Lambda$}{rho\_Lambda}}\label{sec:log}

To solve the Friedmann equations with the running couplings in \eqref{eq:ss_quadratic}, it is convenient to express $G$ in the following form
\begin{align}
G^{-1}(\mu) = G_{0}^{-1}\left( \frac{\mu}{\bar{\mu}} \right)^{2\nu} = G_{0}^{-1}\left(1 + \nu \ln\frac{\mu^{2}}{\bar{\mu}^{2}}\right) + O(\nu^{2})\,.\label{eq:G_exact}
\end{align}
By inserting into \eqref{eq:viablerun}, we find the respective expression for $\rho_\Lambda$ in the form
\begin{align}
\rho_{\Lambda}(\mu) = \rho_{\Lambda 0} + \frac{3\nu}{8\pi (1+\nu)}\big(\mu^{2}G^{-1} - \bar{\mu}^{2}G_{0}^{-1}\big)\,.\label{eq:rho_L_exact}
\end{align}
It is important to stress that Eq.~\eqref{eq:G_exact} does not represent the exact solution of the RG equations. However, both \eqref{eq:G_exact} and \eqref{eq:rho_L_exact} satisfy the conservation law \eqref{eq:cont} for all orders of $\nu$ and, when expanded to first order, reproduce $G$ and $\rho_\Lambda$ in \eqref{eq:ss_quadratic}.

Considering the equation of state $p=w\rho$, the Eq.~\eqref{eq:cont_back} yields $\rho = \rho_0 a^{-3(1+w)}$, where $\rho_0$ is an integration constant. Using this relationship together with \eqref{eq:G_exact}, \eqref{eq:rho_L_exact} and $\mu = H$, one can solve Eq.~\eqref{eq:fried1} for $H(a)$, yielding
\begin{align}
H(a) = H_0 \left\{ \bar{H}_{0}^{2\nu} \left[ (1+\nu)\left(\Omega_{\Lambda0} + \frac{\Omega_0}{a^{3(1+w)}} \right)- \nu\bar{H}_0^{2} \right] \right\}^{\frac{1}{2(1+\nu)}}\,,\label{eq:H_Gln}
\end{align}
where $H_0$ is the Hubble parameter at $a=1$, $\bar{H}_0\equiv \bar{\mu}/H_0$, $\Omega_{0}\equiv 8\pi G_0 \rho_0/3H_{0}^{2}$ and $\Omega_{\Lambda0}\equiv \Lambda_{0}/3H_{0}^{2}$.

To provide a concrete example, it is necessary to specify the value of the reference scale $\bar{\mu}$. Following the approach in Ref.~\cite{Bertini:2024onw}, we can assume that $\bar\mu$ represents the value of the Hubble parameter at the asymptotic de Sitter phase, i.e.,
\begin{align}
\bar{\mu} = \sqrt{\frac{\Lambda_0}{3}} = H_0 \sqrt{\Omega_{\Lambda 0}}\,.
\end{align}
This means that $\bar{H}_0^2 = \Omega_{\Lambda 0}$ and we can then algebraically express $\Omega_0$ as a function of $\Omega_{\Lambda 0}$. From Eq.~\eqref{eq:H_Gln} for $a=1$, we get
\begin{align}
\Omega_0 = - \Omega_{\Lambda0} + \frac{1+ \nu\, \Omega_{\Lambda0}^{1+\nu}}{\Omega^{\nu}_{\Lambda0} \,(1+\nu) } \,.
\end{align}
Expanding up to the first order in $\nu$, $H$ can be written in the form
\begin{align}
H = H_0 F\left[ 1 - \frac{\nu}{2}\left( \frac{a^{-3(1+w)}\ln \Omega_{\Lambda0}}{F^{2}} + \ln\frac{F^{2}}{\Omega_{\Lambda0}} \right) \right]\,, \label{eq:sol-log}
\end{align}
with
\begin{align}
F^{2} = \Omega_{\Lambda0} + (1-\Omega_{\Lambda0})a^{-3(1+w)}\,.
\end{align}

It is important to note that this solution is very similar to the one obtained in \cite{Bertini:2024onw}, differing only by the numerical factor that multiplies $\nu$. Therefore, the solution presented above should lead to similar phenomenologies. In particular, these RG correction may alleviate the tension that exists between local estimates of the Hubble constant and measurements derived from the cosmic microwave background \cite{Abdalla:2022yfr}.

\subsection{Quadratic running of \texorpdfstring{$G$}{G} and quartic running of \texorpdfstring{$\rho_\Lambda$}{rho\_Lambda}} \label{sec:quartic}

In the running of $G$ and $\rho_\Lambda$ in \eqref{eq:ss_quartic1} there is no a priori restriction on the value of $\zeta$, allowing for both $\zeta > 0$ and $\zeta < 0$ \footnote{Similarly, there are no restrictions on $\nu$ in \eqref{eq:ss_quadratic}. However, variations in $\nu$ --- whether positive or negative --- do not lead to significantly different results that merit separate analysis.}. Below, we analyze both cases. For this purpose, it is convenient to make the substitution $\zeta \rightarrow \pm \zeta$ and the following redefinition of constants
\begin{align}
G_\pm = \frac{G_{0}}{1\mp \zeta G_0 \bar{\mu}^{2}}, \quad \rho_{\Lambda \pm} = \rho_{\Lambda 0} \mp \frac{3 \zeta }{16\pi}\bar{\mu}^{4}. \label{eq:Grho}
\end{align}
Here, $G_{+}$ and $\rho_{\Lambda +}$ correspond to the case where $\zeta > 0$, while $G_{-}$ and $\rho_{\Lambda -}$ correspond to $\zeta < 0$. Using this notation, Eqs.~\eqref{eq:fried1} and \eqref{eq:fried2} can be rewritten in the form
\begin{align}
H^{2} \pm \frac{1}{2} \zeta G_\pm H^{4}  = \frac{8\pi G_\pm}{3}\varrho \,, \label{eq:fried12}
\end{align}
\begin{align}
\dot{H}\big(1 \pm \zeta G_\pm H^{2}  \big)  = -4\pi G_\pm (p+\rho)\,,\label{eq:fried22}
\end{align}
where $\varrho = \rho +\rho_{\Lambda\pm}$, and the last equation was obtained by combining \eqref{eq:fried1} and \eqref{eq:fried2} to eliminate $\rho_\Lambda$. It is easy to show that these equations imply \eqref{eq:cont_back}.
\begin{itemize}

\item[$\bullet$] Case $\zeta>0$: By selecting the positive sign in \eqref{eq:fried12} and solving for $H^{2}$, we get
\begin{align}
H^{2}  = -\frac{1}{\zeta G_{+}}\left( 1 \pm \sqrt{1+ \frac{16\pi G_{+}^{2}\zeta}{3} \varrho } \right).
\end{align}
Requiring the above relation to be reduced to the standard Friedmann equation in the limit of low densities, only the sign ($-$) must be retained. The resulting expression has a solution for all $0\leq\varrho<\infty$ provided that $G_{+}>0$; this occurs if $\zeta G_0 H_{0}^{2}<1$, as can be seen from \eqref{eq:Grho}. On the other hand, if $\zeta G_0 H_{0}^{2}>1$ we have $G_{+}<0$, but in this case there is no GR limit.

\item[$\bullet$] Case $\zeta<0$: Choosing the negative sign in \eqref{eq:fried12}, and solving  for $H^{2}$ we find 
\begin{align}
H^{2} = \frac{1}{\zeta G_{-}}\left( 1\pm \sqrt{1-\frac{16\pi G_{-}^{2}\zeta}{3}\varrho} \right).
\end{align}
Imposing that this expression reduces to the standard Friedmann equation in the low density limit, leads us to retain only the ($-$) sign. Then, the physical range of the Hubble parameter lies between zero and the maximum value $H_\mathrm{max} = 1/\sqrt{\zeta G_{-}}$, corresponding to the maximum energy density value $\varrho_\mathrm{max} = 3/(16\pi G_{-}^{2}\zeta)$. Assuming there is no violation of the weak energy condition, $\rho>0$ and $\rho+p \geq 0$, the Eq. \eqref{eq:fried12} implies that the expansion rate will be a monotonously decreasing function of time. 		
\end{itemize} 

The modifications of the Friedmann equations that appear in Eqs.~\eqref{eq:fried12} and \eqref{eq:fried22} have not yet been explored in the context of RG corrections. However, similar modifications arise in other contexts (for example \cite{Barvinsky:2008rd, Lidsey:2009xz, Gao:2012fd, Viaggiu:2015vhs, Bilic:2018ffh, Bertini:2020bar}), and as noted in \cite{Gao:2012fd}, in the modified cosmology described by these equations (with minus sign) the Big Bang singularity is avoided.

\section{Scalar perturbations}\label{sec:pert}

In this section we analyze scalar perturbations in the framework developed in Sec.~\ref{sec:2}. Then, introducing the perturbations in the metric $g_{\alpha\beta} \rightarrow g_{\alpha\beta} + \delta g_{\alpha\beta} $, the linearized Eq.~\eqref{eq:FE} can be written in the form
\begin{align}
\delta G^{\alpha}_{\,\,\,\beta} = 8\pi G(\mu) \left( \frac{d\rho_\Lambda}{d\mu}\delta^{\alpha}_{\beta} - \frac{1}{8\pi}\frac{dG^{-1}}{d\mu}G^{\alpha}_{\,\,\,\beta} \right)\delta\mu + 8\pi G(\mu)\delta T^{\alpha}_{\,\,\,\beta},\label{eq:FE-linear}
\end{align}
where $\delta \mu$ is
\begin{align}
\delta \mu = \frac{1}{3\mu}\delta u^{\alpha} u^{\beta} G_{\alpha\beta} + \frac{1}{6\mu} u^{\alpha} u^{\beta}\delta G_{\alpha\beta},
\end{align}
which is derived by inserting the metric and 4-velocity perturbations into \eqref{eq:ss}.

Assuming a spatially flat background with line element \eqref{eq:metricfried}, we introduce the perturbed line element in the Newtonian gauge
\begin{align}
ds^{2} = (1+2\Psi)dt^{2} - a^{2}(t)(1-2\Phi)\delta_{ij}dx^{i}dx^{j}.
\end{align}
With this $u^{\alpha} = (1-\Psi,\vartheta^{i})$, and consequently
\begin{align}
\delta\mu = \frac{1}{3H}\left[ \frac{1}{a^{2}}\nabla^{2}\Phi - 3H(\dot\Phi + H\Psi)   \right].
\end{align}

We can now go to the perturbed equations of motion. Inserting the above metric components in the field equations \eqref{eq:FE-linear}, we have
\begin{align}
\frac{2}{a^{2}}\nabla^{2}\Phi - 6H( \dot{\Phi} + H\Psi) = 8\pi G(H) \delta T^{0}_{\,\,\,0}\,,\label{eq:lin-00}
\end{align}
\begin{align}
\partial_{i}(\dot{\Phi} + H\Psi) = 4\pi G(H)\delta T^{0}_{\,\,\,i}\,, \label{eq:lin-0i}
\end{align}
\begin{align}
\bigg[\ddot{\Phi} + H(3\dot{\Phi}+\dot{\Psi}) + (3H^{2}+2\dot{H})\Psi - \frac{1}{2a^{2}}\nabla^{2}(\Phi-\Psi)\bigg]\delta^{i}_{j} + \frac{1}{2a^{2}} \partial^{i}\partial_{j}(\Phi - \Psi)
\nonumber\\
+ \frac{1}{3 H}\frac{\dot{G}}{G}\left[ \frac{1}{a^{2}}\nabla^{2}\Phi - 3H(\dot\Phi + H\Psi)   \right]\delta^{i}_{j} = -4\pi G(H)\delta T^{i}_{\,\,\, j}. \label{eq:lin-ij}
\end{align}
The perturbed energy-momentum tensor components are found from Eq.~\eqref{eq:EMT}. Hence, up to the first perturbative order,
\begin{align}
\delta T^{0}_{\,\,\,0} = \delta\rho,\qquad \delta T^{0}_{\,\,\, i} =  -a^{2}(\rho + p)\delta_{ij}\vartheta^{j}, \qquad  \delta T^{i}_{\,\,\, j} = -\delta^{i}_{j}\delta p.
\end{align}
where $\delta\rho$ and $\delta p$ are the energy density and pressure perturbations, respectively.
Therefore, from the off-diagonal part of Eq.~\eqref{eq:lin-ij} one infers the slip parameter\footnote{The slip parameter is defined in Fourier $\boldsymbol{k}$-momentum space but, following the common practice
\cite{Amendola:2016saw}, we omit the dependence on $\boldsymbol{k}$ in Eq.~\eqref{eq:slip}.}
\begin{align}
\eta \equiv \frac{\Phi}{\Psi} = 1, \label{eq:slip}
\end{align}
as in standard GR. Using this result, Eqs. \eqref{eq:lin-00}-\eqref{eq:lin-ij} can be written in Fourier space as
\begin{align}
\frac{k^{2}}{a^{2}}\Phi + 3H( \dot{\Phi} + H\Phi) = - 4\pi G(H) \delta\rho\,,\label{eq:lin-002}
\end{align}
\begin{align}
k^{2}(\dot{\Phi} + H\Phi) =  4\pi  G(H)a^{2}(\rho+p)\theta\,, \label{eq:lin-0i2}
\end{align}
\begin{align}
\ddot{\Phi} + 4H\dot{\Phi} + (3H^{2}+2\dot{H})\Phi  = 4\pi G(H)\bigg( \delta p -  \frac{1}{3 H}\frac{\dot{G}}{G} \delta\rho \bigg) . \label{eq:lin-ij2}
\end{align}
where we define $\theta\equiv \partial_{i}\vartheta^{i}$.

From here we will consider adiabatic perturbations, then the pressure fluctuations can be expressed by $\delta p = c_s^{2}\delta\rho $, where $c_s^{2}\equiv \partial p/\partial\rho$ is the square of the speed of sound. Combining the Eqs.~\eqref{eq:lin-002} and \eqref{eq:lin-ij2}, and using the conformal time $\eta = \int dt/a$, we obtain the closed form equation for the gravitational potential
\begin{align}
\Phi''+\left[ 3(1+c_s^2){\cal H} - \frac{G'}{G} \right]\Phi' + \left[ {\cal H}^{2} +2{\cal H}'  + \left(c_s^{2} - \frac{1}{3{\cal H}}\frac{G'}{G}\right) \left( 3{\cal H}^{2} + k^{2} \right) \right]\Phi  = 0 \, , \label{eq:phi_eta}
\end{align}
where the prime denotes a derivative with respect to $\eta$ and ${\cal H} \equiv a'/a = a H$. 

In the following sections, asymptotic solutions to the above equation are obtained for both short and long wavelength perturbations. This is done using a procedure similar to that in Refs.~\cite{Mukhanov:2005sc, Amendola:2015ksp}, adapted to account for scale-dependent cosmology. To this end, it is convenient to make the change
\begin{align}
\xi = \Phi\exp\left\{ \int \left(\frac{3}{2}(1+c_s^2){\cal H} - \frac{1}{2}\frac{G'}{G}\right) d\eta   \right\} = \frac{\Phi}{\sqrt{G (\rho+p)}},\label{eq:xi}
\end{align}
where we use $c_{s}^{2} = p'/\rho'$ and the continuity equation $\rho'+3{\cal H}(\rho+p) = 0$. Then, using the background equations
\begin{align}
3{\cal H}^{2}  & = 8\pi a^{2}G(\rho + \rho_\Lambda), \label{eq:fried1_eta}\\
{\cal H}' - {\cal H}^{2} & = -4\pi a^{2} G(\rho+p),\label{eq:fried2_eta}\\
\rho'_{\Lambda} & = - \frac{3{\cal H}^{2}}{8\pi a^{2}}\frac{G'}{G^{2}},
\end{align}
the equation for $\xi$ can be written in the form 
\begin{align}
\xi'' - \frac{\sigma''}{\sigma}\xi + \left(c_s^{2} - \frac{1}{3{\cal H}}\frac{G'}{G}\right)k^{2}\xi = 0,\label{eq:diff}
\end{align}
where
\begin{align}
\sigma \equiv \frac{1}{a}\left( \frac{\rho+p}{\rho + \rho_\Lambda} \right)^{-1/2}.
\end{align}

Now that we have derived the perturbed equations, let us move on to the solutions, starting with the large-scale limit.

\subsection{Super-horizon scales}
	
In this case $k\ll {\cal H}$, then the last term of Eq.~\eqref{eq:diff} can be neglected. This leaves us with
\begin{align}
\xi'' - \frac{\sigma''}{\sigma}\xi = 0.\label{eq:dif_sup}
\end{align}
Using the equations for the background \eqref{eq:fried1_eta} and \eqref{eq:fried2_eta}, $\sigma$ can be written in the form
\begin{align}
\sigma = \frac{1}{a}\left(\frac{2}{3}\left( 1- \frac{{\cal H}'}{{\cal H}^{2}}  \right) \right)^{-1/2},\label{eq:sigma}
\end{align}
which is the same expression obtained for standard GR \cite{Mukhanov:2005sc}. Clearly $\xi_1 \propto \sigma$ is one solution of \eqref{eq:dif_sup}, and the other can be found as follows. Let $\xi_2 = v\xi_1$. Substituting in \eqref{eq:dif_sup} one gets $v''\sigma + 2v'\sigma'=0$, whose integration gives $v = c_1(k)\sigma + c_2(k) \sigma \int\sigma^{-2}d\eta$, where $c_1$ and $c_2$ are integrations constants. Thus, the general solution of \eqref{eq:dif_sup} is given by the linear combination of $\xi_1$ and $\xi_2$ in the form \cite{Mukhanov:2005sc}
\begin{align}
\xi = C_1(k) \sigma + C_2(k) \sigma\int_{\eta_i}^{\eta} \frac{d\overline{\eta}}{\sigma^{2}(\overline{\eta})},
\end{align}
where $C_1$ and $C_2$ are integration constants. Using \eqref{eq:sigma} and \eqref{eq:xi}, the solution can be expressed in terms of $\Phi$ as follows
\begin{align}
\Phi & = \overline{C}_1(k) + \overline{C}_2(k)\sqrt{G(\rho+\rho_\Lambda)}\frac{1}{a}\int_{\overline{\eta}_i}^{\eta} a^{2}(\overline{\eta})d\overline{\eta}\,.
\end{align}
The last expression can be rewritten in terms of ${\cal H}$ and $a$ using Eq.~\eqref{eq:fried1_eta}. In this scenario, $\Phi$ retains the same form as in standard GR \cite{Mukhanov:2005sc}, except that here ${\cal H} = {\cal H}(\eta;\nu)$.

\subsection{Sub-horizon scales}

For the case where $k \gg {\cal H}$, the second term in \eqref{eq:diff} can be neglected, then $\xi$ satisfies the equation
\begin{align}
\xi''+ \overline{c}_{s}^{2}k^{2}\xi = 0\,,\label{eq:xi_big_k}
\end{align}
where 
\begin{align}
\overline{c}_{s}^{2}\equiv c_s^2 -\frac{1}{3{\cal H}}\frac{G'}{G}\,.\label{eq:barcs2}
\end{align}
Assuming that $\overline{c}_s^2$ varies slowly, Eq.~\eqref{eq:xi_big_k} can be solved using the WKB approximation, which provides
\begin{align}
\xi \simeq \frac{C(k)}{\sqrt{\overline{c}_s}}\mathrm{exp}\Bigg\{\pm ik\int\overline{c}_s d\eta   \Bigg\},
\end{align}
where $C$ represents the integration constant (one for each signal). As before, this solution has the same shape as the one obtained in standard GR, but here there are corrections to the speed of sound in the fluid. 

Let us continue the analysis by evaluating how this change affects the dynamics of matter. Using $p=w\rho$ (with constant $w$), the perturbed continuity equation $\delta(u^{\beta}\nabla_{\alpha}T^{\alpha}_{\,\,\,\beta}) =0$ is the conventional one
\begin{align}
\delta_\rho'+ 3{\cal H}(c_s^{2}-w)\delta_\rho = -(1+w)(\theta - 3\Phi'), \label{eq:contrast}
\end{align}
where $\delta_\rho \equiv \delta\rho/\rho$ is the density contrast. On the other hand, combining \eqref{eq:lin-0i2} and \eqref{eq:lin-ij2} we have
\begin{align}
\theta' - 3{\cal H}\overline{c}_s^2 \theta + {\cal H}\theta = k^{2}\left( \frac{\overline{c}_s^2}{1+w}\delta_\rho + \Phi  \right). \label{eq:theta}
\end{align}

Now let us evaluate these equations on sub-horizon scales ($k \gg {\cal H}$), with $w=0$ and $\overline{c}_s^2 \ll 1$. In this regime $3{\cal H}\overline{c}_s^2 \theta$ is suppressed with respect to the other terms in \eqref{eq:theta}, just as $3{\cal H}c_s^{2}\delta_\rho$ is suppressed in \eqref{eq:contrast}. In addition, $\Phi' \simeq - {\cal H}\Phi$, and the equation \eqref{eq:lin-002} yields
\begin{align}
k^{2}\Phi = -4\pi G a^{2}\rho\delta_\rho = - \frac{3}{2}{\cal H}^{2}\frac{1}{1+ (\rho_\Lambda/\rho)}\delta_\rho\,.
\end{align}
Deriving and inserting in \eqref{eq:contrast}, we get
\begin{align}
\delta_\rho' = -\theta  - \frac{9}{2}\frac{{\cal H}^{2}}{k^{2}}\frac{\delta_\rho}{1+\rho_\Lambda/\rho}\left( 2\frac{{\cal H}' }{{\cal H}} + \frac{\delta_\rho'}{\delta_\rho} - \frac{(\rho_\Lambda/\rho)'}{1+\rho_\Lambda/\rho}   \right) \simeq - \theta\,.
\end{align}
Consequently, equation \eqref{eq:theta} becomes
\begin{align}
\theta'+{\cal H}\theta = k^{2}\overline{c}_{s}^{2}\delta_\rho - 4\pi G a^{2} \rho \delta_\rho\,.
\end{align}
Combining these last two equations we obtain a second-order differential equation that governs the evolution of the density contrast
\begin{align}
\delta_\rho'' +{\cal H}\delta_\rho' + (k^{2}\overline{c}_{s}^{2} - 4\pi a^{2} G\rho)\delta_\rho = 0 \,.
\end{align}
We can then infer the Jeans length
\begin{align}
\lambda_{\mathrm{J}} \equiv \frac{2\pi a}{k_\mathrm{J}} = \sqrt{\frac{\pi \overline{c}_s^2}{G\rho}}.
\end{align}

Considering the logarithmic running for $G$, and using \eqref{eq:barcs2}, the above expression can be written as
\begin{align}
\lambda_\mathrm{J} = \sqrt{ \frac{\pi c_s^2}{\rho G_0}\left[ 1+\nu \left( \ln\frac{H^{2}}{\bar\mu^{2}} + \frac{1}{c_s^2}\frac{2\dot{H}}{3H^{2}}  \right) \right] }.
\end{align}
The last term in parentheses is dominant and smaller than zero. Therefore, for $\nu > 0$ the effects of RG reduce $\lambda_\mathrm{J}$ and increase the collapse of structures on smaller scales (similar to \cite{Rodrigues:2009vf, Toniato:2017wmk, Bertini:2019xws}). The opposite effect occurs if $\nu < 0$.

Taking into account the quadratic running for $G$ in Eq.~\eqref{eq:ss_quartic1}, then
\begin{align}
\lambda_\mathrm{J} = \sqrt{ \frac{\pi c_s^2}{\rho G_0}\Bigg[ 1+\frac{\zeta}{3}G_0 \Bigg( 3 (H^{2} - \bar{\mu}^{2}) + \frac{2}{c_s^2}\dot{H}  \Bigg) \Bigg] }.
\end{align}
Similarly to the previous case, the last term in parentheses is dominant and negative, thus $\zeta > 0$ implies an increase in the collapse of structures on smaller scales, while $\zeta < 0$ decreases it. However, as observed in \cite{Shapiro:2004ch}, the effects expected for this case are much smaller due to the smallness of $G_0 H^{2}$ ($\dot{H}\propto - H^{2}$ by Eqs.~(\ref{eq:fried1}, \ref{eq:fried2})).

\section{Conclusions}\label{sec:conclusions}

In this work, we consider that the parameters $G$ and $\rho_\Lambda$ vary with the energy scale $\mu$, to which we assign a covariant expression. Using the RG equations, we derive two running models previously obtained in the background cosmology context. However, by adopting the approach developed here, these models can be applied to other contexts as well. To exemplify, scalar cosmological perturbations were developed, and the expression for $\mu$ was derived in the appendix considering a static and spherically symmetric spacetime.

Solutions for background cosmology were obtained by considering two aforementioned specific realizations of $G(\mu)$ and $\rho_\Lambda(\mu)$. The logarithmic running considered in Sec.~\ref{sec:log} has a rich phenomenology on large scales and has been explored before, for example in \cite{Arboleda:2018qdo, Sonia:2022cif, Moreno-Pulido:2022upl, Viaggiu:2021jib} (see also \cite{Toniato:2017wmk, Oliveira:2023mky} for results in other contexts). It is important to emphasize here that the solution \eqref{eq:sol-log} has the same form as that obtained in \cite{Bertini:2024onw} (despite different running models being adopted) and has potential implications for addressing the Hubble tension (e.g., \cite{Abdalla:2022yfr}).  On the other hand, the model considered in Sec.~\ref{sec:quartic} was explored here for the first time and implies that the initial singularity is avoided in a similar way as occurs in the context of holographic cosmology (e.g., \cite{Bilic:2018ffh, Bilic:2015uol, Bertini:2020dvv}) and some approaches that consider higher derivatives \cite{Gao:2012fd, Bilic:2018ffh, Bertini:2020bar}.

Solutions for the scalar modes of the perturbations were derived for any functions $G(H)$ and $\rho_\Lambda(H)$ compatible with the energy-momentum tensor conservation. Cosmological perturbations taking into account the scale dependence at the field equations level  were previously developed in \cite{Fabris:2006gt, Grande:2010vg}, but different expressions for $\mu$ were considered. A feature of the scenario developed here is that explicit modifications appear in the form of an effective sound speed. Clearly, this proposal needs further investigation, not only in late cosmology but also in the early phases of the universe, a period during which the running model in Eq. \eqref{eq:ss_quartic1} is expected to have more significant effects. A covariant expression for $\mu$ also opens up the possibility of developing a complete post-Newtonian expansion. These possibilities constitute future and ongoing work.

\section*{Acknowledgements}

I am grateful to Ilya Shapiro for discussions and comments on this work. The work was supported by {\it Conselho Nacional de Desenvolvimento Científico e Tecnológico} (CNPq -Brazil).

\appendix
\section{Static and spherically symmetrical systems}\label{sec:AP1}

This appendix is dedicated to the development of the scale identification performed in Sec. \ref{sec:2} for applications to systems whose gravitational field is static and spherically symmetric. In this case, the line element can be written in the form \cite{Weinberg:1972kfs}
\begin{align}
ds^{2} = B(r)dt^{2} - A(r)dr^{2} -r^{2}d\theta^{2}-r^{2}\sin^{2}\theta d\varphi\,.
\end{align}
Considering that the components of $u^{\alpha}$ are compatible with a perfect fluid ($u^i=0$), the condition $u^{\alpha}u^{\beta}g_{\alpha\beta}=1$ implies
\begin{align}
(u^{0})^{2} = \frac{1}{B}\,.
\end{align}
Consequently, the expression \eqref{eq:ss} becomes
\begin{align}\label{eq:spherically-mu}
\mu^{2} = -	\frac{1}{3 r^{2}}\frac{d}{dr}\big(rA^{-1}\big)\,.
\end{align}
Next, we need to solve the equations of motion and find $A$. Taking the trace of Eq.~\eqref{eq:FE} to eliminate $R$, we get
\begin{align}
R_{\alpha\beta} = 8\pi G(\mu)\left( T_{\alpha\beta} - \frac{1}{2}g_{\alpha\beta}T - g_{\alpha\beta }\rho_{\Lambda} \right)\,.
\end{align}
For simplicity, we will consider solutions outside the central spherical body, i.e., for $T_{\mu\nu} = 0$. Additionally, on astrophysical scales, the effects of the cosmological constant (and its possible variations) can safely be neglected (see, e.g., \cite{Sereno:2006re}). Thus, the above equation becomes
\begin{align}
R_{\alpha\beta} = 0\,.
\end{align}
The solutions are \cite{Weinberg:1972kfs}
\begin{align}
B(r) = 1 - \frac{k_1}{r} \,, \qquad A = B^{-1}. 
\end{align}
where $k_1$ is a integration constant. With this, Eq.~\eqref{eq:spherically-mu} yields
\begin{align}
\mu^{2} = -\frac{1}{3r^{2}}\,.
\end{align}
Therefore, the dependency $\mu\propto r^{-1}$ is obtained.

\leading{15pt}

\end{document}